\documentclass[
reprint,
superscriptaddress,
showpacs,
 amsmath,amssymb,
 aps,
prx
]{revtex4-1}

\usepackage{graphicx}
\usepackage[ colorlinks,
    linkcolor={blue},
    citecolor={blue},
    urlcolor={blue}]{hyperref}
\usepackage{braket}
\usepackage{dsfont}
\usepackage[varg]{txfonts}

\newcommand\eqt{\hspace{0.17em}{=}\hspace{0.17em}}
\newcommand\equivt{\hspace{0.17em}{\equiv}\hspace{0.17em}}
\newcommand\apt{\hspace{0.17em}{\approx}\hspace{0.17em}}

\newcommand\pt{\hspace{0.17em}{+}\hspace{0.17em}}
\newcommand\mt{\hspace{0.17em}{-}\hspace{0.17em}}
\newcommand\sd{\hspace{0.05em}}
\newcommand\kt{\hspace{0.17em}{<}\hspace{0.17em}}

\newcommand\simt{\hspace{0.17em}{\sim}\hspace{0.17em}}
\newcommand\neqt{\hspace{0.17em}{\neq}\hspace{0.17em}}

\newcommand{\gf}{\hat{G}}
\newcommand{\ham}{\hat{H}}
\newcommand{\ch}{\hat{c}}
\newcommand{\dhat}{\hat{d}}
\newcommand{\se}{\hat{\Sigma}}
\newcommand{\pot}{\hat{V}}

\newcommand{\hc}{^{\dagger}}             %%% suffix for the hermitian conjugate

\newcommand{\mathcomma}{\,,}  %%% comma in mathematical environment
\newcommand{\mathperiod}{\,.} %%% end of sentence in mathematical environment
\newcommand{\nablagrad}{\VV \nabla}   %%% gradient with nabla operator
\newcommand{\varopd}[2][]{\ifthenelse{\equal{#1}{\empty}}{\dd#2}{\dd^{#1}#2}} %%% d for integrals without space, use like \opd{r} or \opd[d]{r}
\newcommand{\vv}[1]{\mathbf{#1}}         %%% vector symbols (std)
\newcommand{\VV}[1]{\boldsymbol{#1}}     %%% vector symbols (for special characters, such as greek letters)

\newcommand{\dd}{\mathrm{d}}             %%% an operator d for total derivatives
\newlength{\mytodolength}

\begin{document}

\title{\textit{Ab initio} spin-flip conductance of hydrogenated graphene nanoribbons: Spin-orbit interaction and scattering with local impurity spins}

\author{Jan Wilhelm}\altaffiliation{Present adress: Department of Chemistry, University of Zurich, Winterthurerstrasse 190, CH-8057 Zurich, Switzerland}\email{jan.wilhelm@chem.uzh.ch}\affiliation{Institute of Nanotechnology, Karlsruhe Institute of Technology, D-76344 Eggenstein-Leopoldshafen, Germany}  \affiliation{Institut f\"ur Theorie der Kondensierten Materie, Karlsruhe Institute of Technology, D-76128 Karlsruhe, Germany}

\author{Michael Walz}\affiliation{Institute of Nanotechnology, Karlsruhe Institute of Technology, D-76344 Eggenstein-Leopoldshafen, Germany} \affiliation{Institut f\"ur Theorie der Kondensierten Materie, Karlsruhe Institute of Technology, D-76128 Karlsruhe, Germany}  
%\affiliation{DFG Center for Functional Nanostructures, Karlsruhe Institute of Technology, D-76131 Karlsruhe, Germany}  

\author{Ferdinand Evers} \affiliation{Institute for Theoretical Physics, University of Regensburg, D-93040 Regensburg, Germany}%\affiliation{Institute of Nanotechnology, Karlsruhe Institute of Technology, D-76344 Eggenstein-Leopoldshafen, Germany} \affiliation{Institut f\"ur Theorie der Kondensierten Materie, Karlsruhe Institute of Technology, D-76128 Karlsruhe, Germany} %\affiliation{DFG Center for Functional Nanostructures, Karlsruhe Institute of Technology, D-76131 Karlsruhe, Germany}  

\vskip 0.25cm

\date{\today}

\begin{abstract}
We calculate the spin-dependent zero-bias conductance $G_{\sigma\sigma'}$ in armchair graphene nanoribbons with hydrogen adsorbates  
employing a DFT-based \textit{ab initio} transport formalism including spin-orbit interaction. 
We find that the spin-flip conductance $G_{\sigma\bar{\sigma}}$ can reach the same order of magnitude as the spin-conserving one, 
$G_{\sigma\sigma}$, due to exchange-mediated spin scattering. In contrast, the genuine spin-orbit interaction appears to 
play a secondary role, only. 
\end{abstract}

\verb



\maketitle

\section{Introduction}
In recent years, graphene~\cite{graphene} has been considered as ideal spintronics material~\cite{ReviewFabian}: 
Due to the weak spin-orbit interaction (SOI)~\cite{GmitraPRB}, spin lifetimes of Dirac electrons are expected to be long. 
However, the %typical
 Hanle precession measurements 
typically 
yield spin diffusion times several orders of magnitude below the 
theoretical predictions~\cite{ReviewRoche}. 
Recently, a quantum interference measurement~\cite{MagnImpDephExp} proposed that intrinsic local magnetic moments at 
defects~\cite{DefMagnet} are the primary cause of spin relaxation in graphene, masking any potential effects of the genuine SOI. 
The efficiency of such a mechanism was confirmed by consecutive  theoretical work~\cite{PRLREGMAGNETICIMP}. 
While another mechanism for spin-flips originating from bias induced orbital magnetism has also been identified, recently, 
its quantitative effect still remains to be explored~\cite{Ringcurrents}.

Motivated by the efficiency of exchange induced spin-flips in graphene, 
we study spin-dependent transport in  graphene nanoribbons (GNRs), 
i.e.~strips of graphene with ultra-thin width, 
from  first principles. Our interest in GNRs is closely related to their electronic 
properties: GNRs inherit a weak intrinsic SOI~\cite{PRBBurkard2} and high electron mobility from graphene~\cite{PRBBurkard1}.
Moreover, GNRs exhibit 
% a bandgap 
gaps 
that can be tuned with the ribbon width~\cite{energygapsgraphenenanoribb,
*Michelnanoribbon}  and  local spins can be generated at zigzag edges~\cite{RibbonTheoFirst,*zigzagDressel,*zigzagLouie,*expzigzag} or defects~\cite{DefMagnet}.
These properties make GNRs promising materials for applications in spintronics, e.g.~for quantum computing~\cite{Quantencomp}.

Spin transport in GNRs will be addressed in this paper using the standard formalism of molecular electronics~\cite{DiVentra}: 
A device scattering region is located between two semi-infinite leads with applied bias~$V$; 
the total conductance $G\eqt dI/dV$ of the device is split into four spin-dependent conductance 
coefficients~$G_{\sigma\sigma'}$ (with $G\eqt\sum_{\sigma\sigma'}G_{\sigma\sigma'}$). They account for an electronic current being injected with spin~$\sigma$,
and, after passing the device region, measured with spin direction $\sigma'$. 

In ribbons with spin-degenerate electronic structure and neglected spin-flip scattering, 
$G$ splits equally, $G_{\uparrow\uparrow}\eqt G/2$ and $G_{\downarrow\downarrow}\eqt G/2$~\cite{CondQuantGr1,*CondQuantGr2,*CondQuantGr3,*CondQuantGr4,*CondQuantGr5,*CondQuantGr6}. 
In the case of magnetic ribbons, $\uparrow$- and $\downarrow$-current do not match anymore; unpaired spins in the ribbon, e.g.~at defects or zigzag edges, 
can cause characteristic differences between $G_{\uparrow\uparrow}$ and $G_{\downarrow\downarrow}$~\cite{zigzag1,*updowntrhydr,*zigzag2,*zigzag3,*zigzag4,*zigzagprobeRegensburgnichtFabian}.

The spin-flip conductance coefficients $G_{\uparrow\downarrow}$ and $G_{\downarrow\uparrow}$ 
are non-vanishing in the presence of  (i) SOI~\cite{GsfPareekBruno,*GsfSOC6,*GsfSOC,*GsfSOC3,*GsfSOC2,*GsfSOC4,*GsfSOC5,*SFCRibbons} 
or (ii) exchange interaction with local spins in the device~\cite{GsfMagnImp}. 
To include %the 
SOI in our DFT formalism, we employ an all-electron SOI module~\cite{SOCMiddendorf}, 
the exchange-interaction is dealt with on the level of spin DFT~\cite{GGA}; details 
see Sec.~\ref{sec:sec2}. 
%%%%%%%%%%%%%%%%%%%%%%%%%%%%%%%%%%%%%%%%%%%%%%%%%%%%%%%%%%%%%%%%%%%%%%%%%%%%%%%%%%%%%%%%%%%%%%%%%%%%%%%%%%%%%%%%%

As one would expect, we find a very small spin-flip conductance~$G_{\sigma\bar{\sigma}}$ in clean armchair GNRs (AGNRs)
 due to the very weak SOI and the absence of local impurity spins, see Sec.~\ref{sec:sec3}. In contrast, the spin-flip 
conductance is massively enhanced  in the presence of adsorbates, 
see Sec.~\ref{sec:sec4}\,-\,\ref{sec:sec7}. For instance, our results indicate that the spin-flip probability associated with 
a single hydrogen adatom can be comparable to the spin-conserving one.  
%%%%
This high spin-flip probability is rationalized by employing a simplistic tight-binding (toy) model. 
% in  the Appendix~\ref{sec:sec5}.  
Our first-principles results are qualitatively similar to analytical results by earlier authors, 
Ref.~\cite{PRLREGMAGNETICIMP}, 
who employ a model calculation that is valid in the 
highly dilute limit.

\section{Method} \label{sec:sec2}
In our calculations, we are employing an extension of the AITRANSS platform, 
our DFT-based transport simulation tool~\cite{Transportcode,*STM,*AlexejTransport,*PaulsPaper,MichaelsMethod}. 
The spin-dependent conductance  is obtained as follows: 
We extract the Kohn-Sham (KS) Hamiltonian $\ham\eqt[(\ham_{\uparrow\uparrow},\ham_{\uparrow\downarrow}),(\ham_{\downarrow\uparrow},\ham_{\downarrow\downarrow})]$, 
a $2 {\times} 2$ block-matrix in spin space, from a DFT calculation~\footnote{Methodological details: TURBOMOLE 
package~\cite{Turbomole}, DFT with generalized gradient 
approximation~(GGA~\cite{GGA}, BP86 functional~\cite{BP86})
together with a contracted  Gaussian-type basis (def2-SVP)~\cite{def2SVP} 
and corresponding Coulomb-fitting basis set within the resolution of the 
identity (RI) approximation~\cite{RI} in a two-component formalism~\cite{B2CFormalism,*B2CFormalism2}.} including all-electron SOI~\cite{SOCMiddendorf} for a 
finite-size hydrogen-terminated graphene nanoribbon 
with horizontal armchair edges, see Fig.~\ref{fig1}.  Subsequently, we obtain  the (retarded) 
single particle KS-Green's function~$\hat{G}$ of a finite-size strip 
in the presence of the left and right contacts by  standard 
recursive Green's function techniques~\cite{DiVentra}: 
\begin{align}
 \hat{G}(E)
\equiv \left(\begin{array}{cc}\gf_{\uparrow\uparrow} &\gf_{\uparrow\downarrow} \\[0.5em]\gf_{\downarrow\uparrow} & \gf_{\downarrow\downarrow}    \end{array}\right)
  = \left( \hat{G}_0^{-1} - \se^L-\se^R\right)^{-1}
 . \label{eq1}
\end{align}
The spin-diagonal self-energies $ \se^\alpha\eqt[(\se^\alpha_\uparrow,0),(0,\se_\downarrow^\alpha)]$ with $\se_\uparrow^\alpha\eqt\se_\downarrow^\alpha$  reflect the 
presence of the leads~\cite{ABC}. They are treated with a closed-shell electronic structure 
and a vanishing SOI so that spin is a good quantum number in the leads~\cite{PSCI}. 
$\hat G_{0}$~represents the bare KS-Green's function of the device region, see Fig.\,\ref{fig1}\,(a). 
We compute the spin-dependent zero-bias conductance $G_{\sigma\sigma'}(E)$ at a given chemical potential~$E$ [see Fig.\,\ref{fig1}\,(b)] in a Landauer-B$\ddot{\text{u}}$ttiker approach~\cite{Land,*Buett}:
\begin{align}
G_{\sigma\sigma'}(E)=\frac{e^2}{h} \,\text{Tr}\hspace{-0.1em}\left[\hat{\Gamma}^L_\sigma \,(\gf_{\sigma\sigma'})^\dagger\,\hat{\Gamma}^R_{\sigma'}\, \gf_{\sigma'\sigma} \right],
\label{eq3}
\end{align}
with $\hat{\Gamma}_\sigma^\alpha\eqt i(\se^\alpha_\sigma \mt (\se^\alpha_\sigma)^\dagger)$.

For spin quantization $\mathbf{n}\eqt(\sin\theta\cos\varphi,\sin\theta\sin\varphi,\cos\theta)$ 
deviating from default $z$-direction, we rotate the Green's function in spin space by the unitary transform~$U$:
 \begin{align}
\gf^{ ({\mathbf{n}})} = U \gf U^\dagger\,,\hspace{0.3cm} U =  \left(\begin{array}{cc} \cos \frac{\theta}{2}&  
-e^{-i\varphi}\sin\frac{\theta}{2} \\[0.5em] e^{i\varphi}\sin\frac{\theta}{2}  &  \cos\frac{\theta}{2}   \end{array}\right) \,.
 \end{align}
Due to the closed-shell electronic structure of the leads, $\hat{\Gamma}_\sigma^\alpha$ remains unchanged by a 
unitary transform and the conductance with respect to an arbitrary spin quantization axis~$\mathbf{n}$ is given by
\vspace{-1em}\begin{align}
G_{\sigma\sigma'}^{ ({\mathbf{n}})}(E)=\frac{e^2}{h} \,\text{Tr}\hspace{-0.1em}\left[\hat{\Gamma}^L_\sigma \,(\gf_{\sigma\sigma'}^{ ({\mathbf{n}})})^\dagger\,\hat{\Gamma}^R_{\sigma'}\, \gf_{\sigma'\sigma}^{ ({\mathbf{n}})} \right].\label{eq5}
\end{align}
The formalism outlined here is well established~\cite{GsfPareekBruno,GsfSOC6,GsfSOC,GsfSOC3,GsfSOC2,GsfSOC4,GsfSOC5,SFCRibbons}. 
%in particular for differing spin quantization of conducting electrons~\cite{GsfPareekBruno} and for GNRs~\cite{SFCRibbons}.

\section{Results} \label{sec:sec3real}
In this section, we present simulation results of the spin-dependent conductance~$G_{\sigma\sigma'}$  of a clean 
and hydrogenated AGNRs calculated according to Eqs.\,\eqref{eq3} and~\eqref{eq5}. For computational details, we refer to App.~\ref{sec:app2}.
%%%%%%%%%%%%%%%%%%%%%%%%%%%%%%%%%%%%%%%%%%%%%%%%%%%%%%%%%%%%%%%%%%%%%%%%%%%%%%%%%%%%%%%%%%%%%%%%%%%%%%%%
\subsection{Clean ribbon}\label{sec:sec3}
%%%%%%%%%%%%%%%%%%%%%%%%%%%%%%%%%%%%%%%%%%%%%%%%%%%%%%%%%%%%%%%%%%%%%%%%%%%%%%%%%%%%%%%%%%%%%%%%%%%%%%%%
\begin{figure}[t]
\centering
\includegraphics[width=8.6cm]{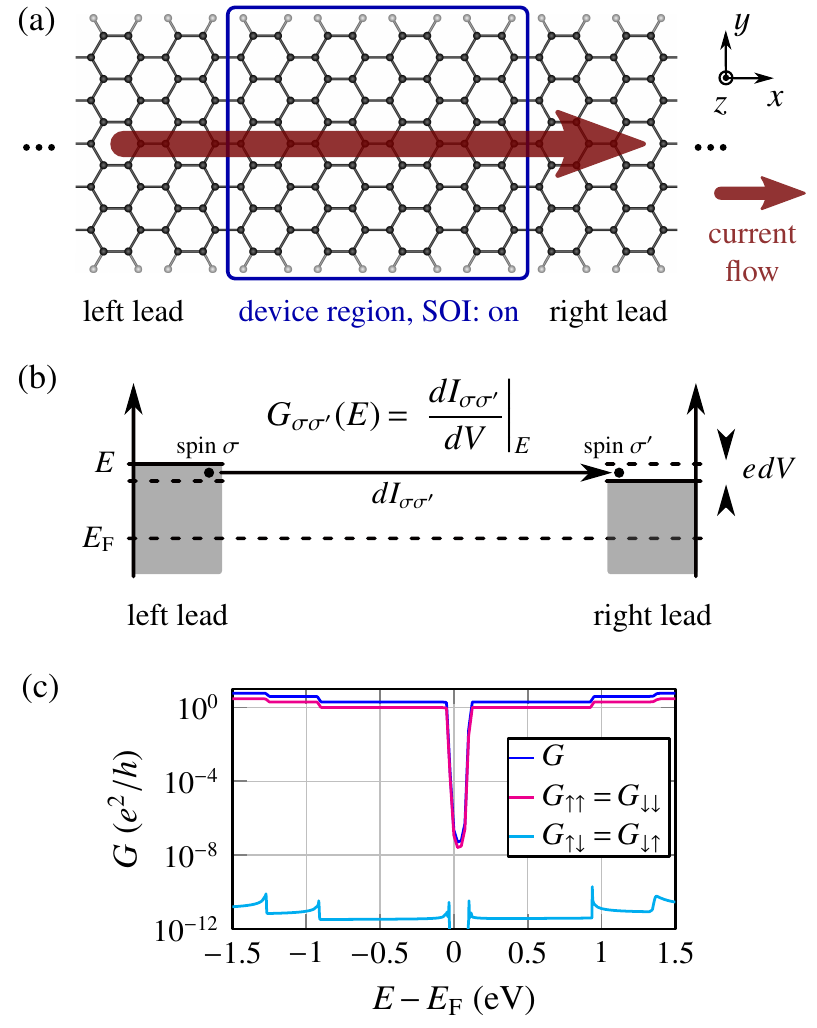}
\caption{(a) Structure of a clean AGNR11 device. In the blue marked device region, SOI is present. 
(b) Between the two leads of (a), an infinitesimal voltage $dV$ is applied. 
As response, an electronic current $dI$ flows, which is splitted into four components $dI_{\sigma\sigma'}$. They account for an electronic current being injected with spin~$\sigma$ and measured after passing the device region with spin~$\sigma'$. The spin-dependent conductance $G_{\sigma\sigma'}(E)$ at a given chemical potential~$E$ is then defined as ratio of $dI_{\sigma\sigma'}$ and $dV$. The reference energy $E_\text{F}$ 
is the chemical potential of the isolated, charge-neutral device.
  (c) Corresponding conductance of the pristine ribbon (a) according to Eq.\,\eqref{eq3}. 
The spin quantization axis is chosen in positive $z$-direction. The spin-conserving conductance 
does not vanish completely inside the bandgap (but is reduced by seven orders of magnitude) due 
to the limited horizontal size of the lead system, so that the one-dimensional band structure 
of the ribbon cannot be formed with ultra-high precision. }
\label{fig1}
\end{figure}
We calculate the conductance coefficients $G_{\sigma\sigma'}(E)$ of an AGNR with $N_\text{C}\eqt11$ transverse carbon 
atoms (AGNR11), see Fig.\,\ref{fig1}\,(a) for the molecular structure of the device and  Fig.\,\ref{fig1}\,(b) for a sketch of the electronic structure and the transport setup. The result for the spin-dependent conductance according to 
Eq.\,\eqref{eq3} is shown in Fig.\,\ref{fig1}\,(c) with a spin quantization axis in $+z$-direction. 
For the spin-conserving conductance coefficients $G_{\sigma\sigma}$, we find a step function with $G_{\sigma\sigma}(E)/(e^2/h)$  
simply counting the energy bands intersecting with a given energy $E$ 
%JW: (within 10$^{-2}$ precision)
~\cite{TransmNumberofBands,*TransmNumberofBands2}.   
$G_{\sigma\sigma}$ is hardly affected by the SOI. Most importantly, the spin-flip conductance is 
found to be very small with an upper bound of $10^{-10}\,e^2/h$ due to the very weak SOI as 
expected.~\footnote{When changing the horizontal length of the device region with SOI, we observe an unchanged spin-flip 
conductance. We conclude that the spin-flip scattering occurs at the lead-device crossover probably caused by exchange with the magnetic zigzag edge states of the finite-size DFT calculation leaking into the device region, see Sec.~\ref{sec:sec4} and App.~\ref{subsec:apppart} for a detailed discussion. This conceptual issue 
also reflects in the linear (and not quadratic) scaling of $G_{\sigma\bar{\sigma}}$ with the DOS, 
$G_{\sigma\bar{\sigma}}(E)\simt\rho(E)$, in the energy interval~$[-\,1.5\,\text{eV},1.5\,\text{eV}]$. 
However, we can restrict the methodogical error of $G_{\sigma\bar{\sigma}}$ due to spin-flip 
scattering at the lead-device crossover by an upper bound of $10^{-10}\,e^2/h$ which is orders 
of magnitude below the spin-flip conductance of hydrogenated AGNRs. Additionally, the spin-flip 
conductance of the clean AGNR11 due to SOI is bounded by $10^{-10}\,e^2/h$.}
Due to vertical mirror symmetry, ${\uparrow}_z$ and ${\downarrow}_z$ bands of pristine AGNRs 
are degenerate~\cite{SOCNanoribbonSymmetry} and therefore 
$G_{\uparrow\uparrow}(E) \eqt G_{\downarrow\downarrow}(E)$ and $G_{\uparrow\downarrow}(E) \eqt  G_{\downarrow\uparrow}(E)$ for $z$-quantization.

%%%%%%%%%%%%%%%%%%%%%%%%%%%%%%%%%%%%%%%%%%%%%%%%%%%%%%%%%%%%%%%%%%%%%%%%%%%%%%%%%%%%%%%%%%%%%%%%%%%%%%%%
\subsection{Ribbon with a single hydrogen adatom}\label{sec:sec4}
\begin{figure}[t]
\centering
\includegraphics[width=8.6cm]{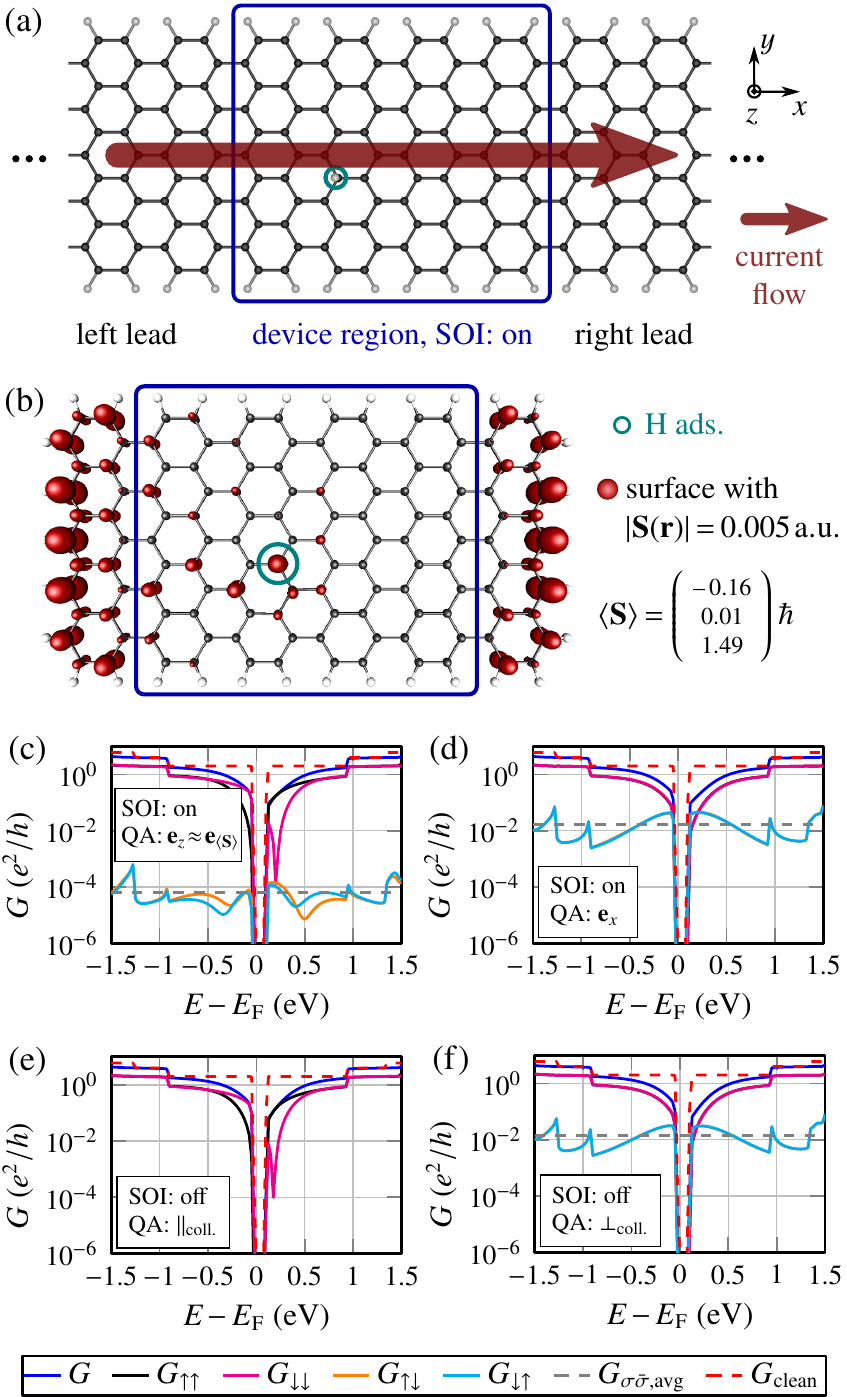}
\caption{(a)  Structure of the AGNR11 device with one single hydrogen adsorbate. 
(b) Finite-size ribbon for the underlying DFT calculation including SOI. 
All atoms in the blue box belong to the device region. The red surface 
denotes the isosurface of the spin density calculated by the DFT including SOI where 
we observe the famous zigzag edge magnetism~\cite{RibbonTheoFirst,zigzagDressel,zigzagLouie,expzigzag}. The total spin is calculated as
 $\braket{\mathbf{S}}{=}\int \mathbf{S}(\mathbf{r}) \sd d^3\mathbf{r}$. (c)\,--\,(f) 
Conductance of device region (a) with/without SOI and quantization axis according to inset. 
The reference energy, $E_\text{F}$ 
is the chemical potential of the isolated, charge-neutral lead. 
The average spin-flip conductance $G_{\sigma\bar{\sigma}}$ is computed as arithmetic mean of 
$G_{\uparrow\downarrow}$ and $G_{\downarrow\uparrow}$ in the energy interval~$[-\,1.5\,\text{eV},1.5\,\text{eV}]$. 
}
\label{fig2}
\end{figure}

%%%%%%%%%%%%%%%%%%%%%%%%%%%%%%%%%%%%%%%%%%%%%%%%%%%%%%%%%%%%%%%%%%%%%%%%%%%%%%%%%%%%%%%%%%%%%%%%%%%%%%%%
We continue with a ribbon containing a single hydrogen adatom, see Fig.\,\ref{fig2}\,(a) 
for the molecular geometry. Hydrogen forms a chemical bond with the underlying carbon 
atom resulting in an sp$^3$ hybridization. The four nearest carbon atoms were structurally 
relaxed in order to catch the massive enhancement of SOI due to the lattice 
distortion~\cite{PRLSOCImpGraphene,*PRLSOCHGraphene,*SOCHydrGrapheneBala,*SOCfluor}. 

First, we comment on the finite-size  DFT calculation  of the ribbon including SOI as sketched in Fig.\,\ref{fig2}\,(b): 
The computed magnetization of the finite-size ribbon  
is ${\braket{\mathbf{S}}}\eqt(-\sd0.16,-\sd0.01,1.49)^T\hbar$~\footnote{In a 
non-SOI DFT calculation, there is no spatial information about the magnetization axis. 
In contrast, we get a spatial direction of the magnetization in a SOI DFT calculation. 
The reason for the magnetization axis  being so close to the $z$-axis is the following: 
The extended H$\ddot{\text{u}}$ckel guess as starting point for the SOI-DFT calculation 
computes one unpaired electron with spin polarization axis in default $z$-direction. 
This magnetization axis  was already a local minimum and the self-consistent-field iteration 
of SOI DFT doesn't deviate much from this starting axis~\cite{BeljNanoLett}. }, 
so there are approximately three unpaired electrons in the ribbon distributed 
at the zigzag edges~\cite{RibbonTheoFirst,zigzagDressel,zigzagLouie} and 
near the impurity, see Fig.\,\ref{fig2}\,(b). 
\footnote{In this work, we do not discuss Kondo physics focusing on temperatures well above 
the Kondo scale $T_\text{K}$. Indeed, there appears to be experimental evidence for a 
relatively small Kondo temperature in hydrogenated graphene, 
$T_\text{K}\hspace{0.12em}{\lesssim}\hspace{0.17em} 2\,\text{K}$~\cite{HatomlocstateexpGrig}. 
For a recent review on Kondo impurities in graphene, we refer to Ref.~\cite{Kondo1}.}

%%%%%%%%%%%%%%%%%%%%%%%%%%%%%%%%%%%%%%%%%%%%%%%%%%%%%%%%%%%%%%%%%%%%%%%%%%%%%%%%%%%%%%%%%%%%%%%%%%%%
Our simulation results for the spin-dependent conductance~$G_{\sigma\sigma'}$ according to 
Eqs.\,\eqref{eq3} and \eqref{eq5} are displayed in Fig.\,\ref{fig2}\,(c)\,--\,(f). 
%showing two different quantization axes and calculations with SOI and without SOI. 
First, we focus on the situation where the spins 
of sample and incoming electrons are (very nearly) aligned. 
The corresponding spin-conserving conductance~$G_{\sigma\sigma}$  is displayed in 
Fig.\,\ref{fig2}\,(c) with SOI 
and (e) without SOI. 
%for incoming electrons with spin axis parallel to $\braket{\mathbf{S}}$. 
The values for 
$G_{\uparrow\uparrow}(E)$ and $G_{\downarrow\downarrow}(E)$  
in  Fig.\,\ref{fig2}\,(c) and (e)  deviate by less than 
10$^{-2}\,e^2/h$, only, and we observe that SOI hardly influences $G_{\sigma\sigma}$. 
In particular, the broad antiresonance~\cite{LiLu,RocheBNTransm,*RocheChemTransm,*RocheOTransm,*ImpItalien,*RocheEdge,*ManyimpuritiesCanada,*RocheBNmany,*Chiu1,*ImpFinnland,*StoneWales1,*StoneWales2,*PRBPaper,*Chiu2,*Petrovic} indicating the quasilocalized state (zero mode) that 
accompanies the isolated adatom remains 
clearly visible also in the presence of SOI.
%%%%
In Fig.\,\ref{fig2}\,(c), the spin-flip 
conductance is seen to be very small, $G_{\sigma{\bar{\sigma}}}(E) \simt 10^{-4}\,e^2/h$,
while it vanishes in Fig.\,\ref{fig2}\,(e) due to the absence of SOI.

%%%%%%%%%%%%%%%%%%%%%%%%%%%%%%%%%%%%%%%%%%%%%%%%%%%%%%%%%%%%%%%%%%%%%%%%%%%%%%%%%%%%%%%%%%%%%%%%%%%%555
Next, we consider in Fig.\,\ref{fig2}\,(d) incoming electrons with spin polarization along $x$-axis,
i.e.~perpendicular to the magnetization axis $\apt \mathbf{e}_z$ of local spins in the ribbon. 
%~\footnote{The conductance of the ribbon in Fig.\,\ref{fig2}\,(a) for incoming electrons being polarized 
%along the $y$-axis matches quantitatively with the one for $x$-polarization.}. 
Here, the spin-flip conductance increases strongly reaching values larger than 0.05\,$e^2/h$. Near the bandgap,
it is even exceeding the spin-conserving conductance. In order to emphasize that the large spin-flip conductance is due to the exchange 
interaction and not related to SOI, we repeat the same calculation without SOI. The result is shown in 
Fig.\,\ref{fig2}\,(f) and indeed it is indistinguishable from Fig.\,\ref{fig2}\,(d). 
In a nutshell, the exchange-driven spin flip is understood as follows~\cite{spsc}: 
Say, the fixed impurity spin points into the $z$-direction, 
i.e.~the exchange interaction turns into 
$S_{\text{imp}}^z\,\hat{{S}}_{\text{cond}}^z\simt\hat{\sigma}_\text{cond}^{\,z}$ 
with the Pauli matrix $\hat{\sigma}_\text{cond}^{\,z}$ acting on incoming conduction electrons.
As a consequence, the effective single-particle Hamiltonian no longer commutes with 
$\hat{\sigma}_\text{cond}^{\,x,y}$. Hence, the spin of the incoming 
particles no longer is conserved, if it happens 
to exhibit a component perpendicular to the impurity spin.
%to point in the $x,y$-plane
Therefore, spin-flips become possible with a probability $(G_{\uparrow\downarrow}\pt G_{\downarrow\uparrow})/\sum_{\sigma\sigma'}G_{\sigma\sigma'}$ that can 
reach order unity for a non-collinear spin passing a single hydrogen adatom. 
%%%
In App.~\ref{sec:sec5}, we explain how our results are rationalized employing a simple 
toy model. Our overall findings are consistent with Ref.~\cite{PRLREGMAGNETICIMP} that has employed a model 
calculation.

%%%%%%%%%%%%%%%%%%%%%%%%%%%%%%%%%%%%%%%%%%%%%%%%%%%%%%%%%%%%%%%%%%%%%%%%%%%%%%%%%%%%%%%%%%%%%%%%%%
\subsection{Ribbon with two hydrogen adatoms}\label{sec:sec6}
%%%%%%%%%%%%%%%%%%%%%%%%%%%%%%%%%%%%%%%%%%%%%%%%%%%%%%%%%%%%%%%%%%%%%%%%%%%%%%%%%%%%%%%%%%%%%%%%%%

\begin{figure}[t]
\centering
\includegraphics[width=8.6cm]{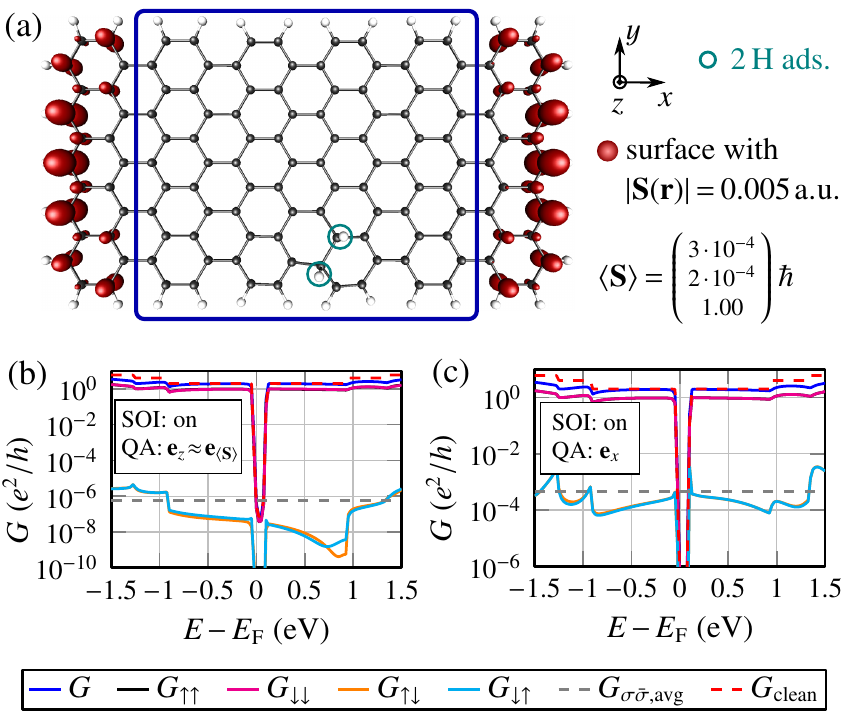}
\caption{(a) Finite-size ribbon with two hydrogen adatoms for the underlying DFT calculation including SOI. 
The atomic structure is relaxed including the six surrounding carbon atoms. 
All atoms in the blue box belong to the device region. (b) Spin-dependent conductance for 
$z$-polarized electrons and (c) for $x$-polarized electrons. }\vspace{-0.2em}
\label{fig5}
\end{figure}

We calculate the spin-dependent conductance of a ribbon with two neighboring 
hydrogen adatoms, see Fig.\,\ref{fig5}\,(a). This double-hydrogen defect 
is non-magnetic and we would like to confront it with the case of an isolated hydrogen adatom. 
In Fig.\,\ref{fig5}\,(a), the non-magnetic character of this impurity is evident: The spin density  
near the impurity is smaller than the isovalue and so it cannot be resolved anymore. 
There are two reasons for the absence of magnetism with this defect. 
%%%%
(i) There is no imbalance between the graphene sublattices, i.e.~one impurity on each sublattice. 
This also implies that the number of electrons remains even.  
%%%
(ii) Because of the vicinity of the border, orbital degeneracies are lifted, so that 
a closed-shell ground state is favored.

The computed conductance  is shown in Fig.\,\ref{fig5}\,(b) for the collinear case and
in Fig.\,\ref{fig5}\,(c) for the non-collinear one. 
Indeed, the spin-flip conductance for non-collinear transport is two orders of 
magnitude smaller as compared to the previous case with a single hydrogen adatom. 
Note, that again, the spin-flip conductance for electrons polarized perpendicular to the sample magnetization 
exceeds the collinear one --  probably due to weak residual magnetism with spin density 
$|{\mathbf{S}(\mathbf{r})}|\kt 0.005\,\text{a.\hspace{0.03em}u.}$.

We conclude that the interactions between hydrogen adatoms can be relevant, if they come 
sufficiently close. In our example, the spin-flip conductance of the ribbon with two hydrogen adatoms is far less 
than twice the value of the isolated adatom. This indicates a breakdown of Matthiesen's rule in the 
limit of higher concentrations.

%%%%%%%%%%%%%%%%%%%%%%%%%%%%%%%%%%%%%%%%%%%%%%%%%%%%%%%%%%%%%%%%%%%%%%%%%%%%%%%%%%%%%%%%%%%%%%%%%%
%%%%%%%%%%%%%%%%%%%%%%%%%%%%%%%%%%%%%%%%%%%%%%%%%%%%%%%%%%%%%%%%%%%%%%%%%%%%%%%%%%%%%%%%%%%%%%%%%%

%%%%%%%%%%%%%%%%%%%%%%%%%%%%%%%%%%%%%%%%%%%%%%%%%%%%%%%%%%%%%%%%%%%%%%%%%%%%%%%%%%%%%%%%%%%%%%%%%%
\subsection{Massively hydrogenated ribbon}\label{sec:sec7}
%%%%%%%%%%%%%%%%%%%%%%%%%%%%%%%%%%%%%%%%%%%%%%%%%%%%%%%%%%%%%%%%%%%%%%%%%%%%%%%%%%%%%%%%%%%%%%%%%%

\begin{figure}[t]
\centering
\includegraphics[width=8.6cm]{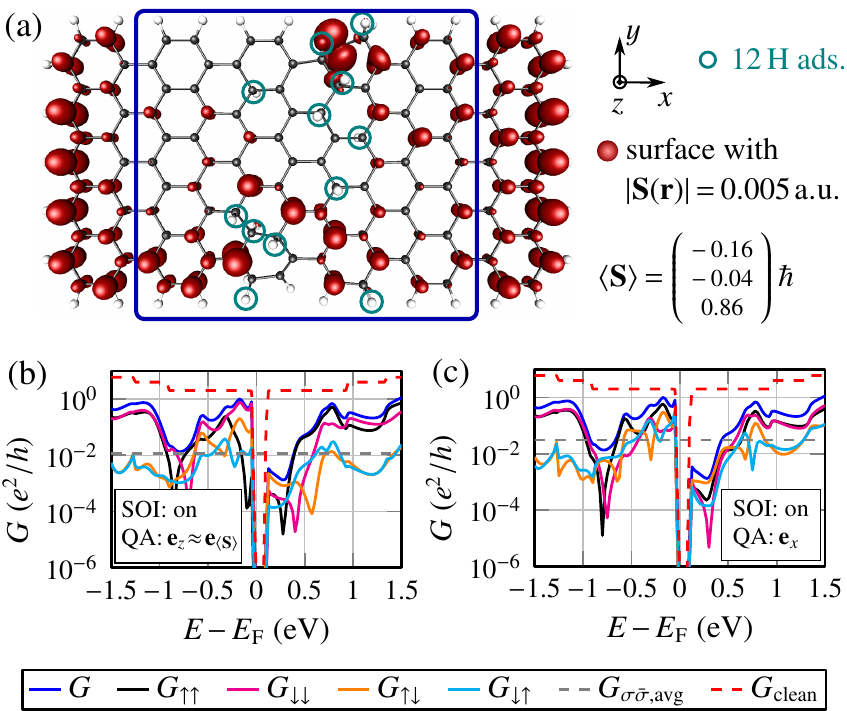}
\caption{(a) Finite-size ribbon with 12 hydrogen adatoms  for the underlying DFT calculation including  SOI. The total number 
of unpaired electrons is $N_\text{Spin}\eqt 2/\hbar\, |\hspace{-0.15em}\braket{\mathbf{S}}\hspace{-0.15em}|\eqt 1.75$. 
All atoms in the blue box belong to the device region. (b) Spin-dependent conductance for electrons polarized 
along $\mathbf{e}_z$ and (c) for $x$-polarization. 
}
\label{fig6}
\end{figure}

In Fig.\,\ref{fig6}, we show the spin-dependent conductance for a ribbon with 12 adsorbed hydrogen 
atoms which are  distributed randomly on the ribbon. The part of the structure that carries 44 carbon atoms and 12 adsorbed hydrogen 
atoms was structurally relaxed. The transmission function displayed in Fig.\,\ref{fig6}\,(b) and (c) 
is seen to carry strong mesoscopic fluctuations that reflect many quasilocalized states near the Fermi energy. 
In this situation, the spin-flip conductance can reach the same order of magnitude as the spin-conserving one in sizable energy window. 

Notice, that even for incoming electrons with collinear spin along the direction of~${\braket{\mathbf{S}}}$, 
the spin-flip conductance is very large. We interpret this effect as an indication that the 
direction of the local spin density is fluctuating in space. 

\subsection{Local current density in massively hydrogenated ribbon}\label{subsec:currentdensityplot}

\begin{figure}[t]
\centering
\includegraphics[width=8.6cm]{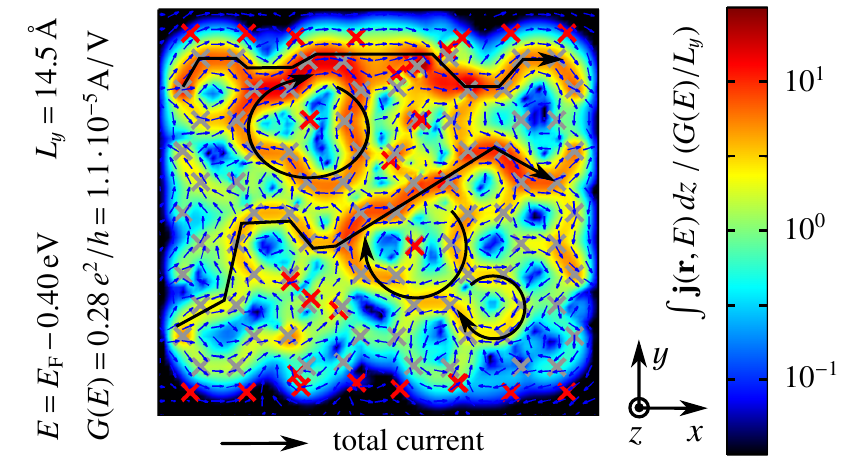}
\caption{
Local current density response (integrated over the out-of-plane direction; normalized to the conductance per width) in the hydrogenated AGNR11 shown in Fig.\,\ref{fig6}\,(a) (blue box only). The current density exhibits strong mesoscopic fluctuations that reflect in a logarithmic color scale covering three decades. Some interesting current paths are drawn into the picture for illustration: Local current vortices exceeding the spatial average current by one order of magnitude (see dark red regions). Plot shows current amplitude (color), current direction (arrows), carbon atoms
(grey crosses) and hydrogen atoms (red crosses). 
}
\label{fig:j}
\end{figure}

In Fig.~\ref{fig:j}, we show the local current density response for the massively hydrogenated AGNR11 of Sec.~\ref{sec:sec7} based on an open-shell DFT 
calculation including SOI (see App.~\ref{sec:currents} for method details; see \,Fig.\,\ref{fig6}\,(a) for the atomic structure). 

The current exhibits strong mesoscopic fluctuations covering three orders of magnitude. They are 
related to vortices which exceed the average current by over one order of magnitude. 
Such current vortices go along with orbital magnetism which are also potentially 
relevant for spin relaxation~\cite{Ringcurrents}.

We compare this finding with our recent work~\cite{Ringcurrents}, in which we simulated the local current density for a larger 
ribbon [$N_\text{C}\eqt41$ transverse carbon atoms (AGNR41)] but enforced a closed-shell electronic structure without SOI to 
reduce the computational effort~\footnote{Apart from a factor of two in the matrix sizes, the main computational cost for the full 
spin treatment is a slow convergence in the self-consistent field (SCF) cycle of the DFT. This is due to many possible spin 
configurations with (nearly) equal energy. Physically, this is the same effect as in a spin glass~\cite{SpinGlass}: flipping a 
single spin is energetically hardly suppressed if neighboring spins are also flipped. The larger the system gets, the more low 
lying spin excitation exist in a given energy window; making the SCF convergence extremely costly for larger systems.}. 
The characteristic features, i.e., current vortices and broad fluctuations, are the same in both cases; they emerge from the 
full spin  treatment including SOI as well as from the spin restricted treatment. Therefore, we 
confirm that their appearance 
is a fundamental property of the scattering states in defected graphene flakes.

\section{Conclusion}
In conclusion, we calculate the spin-dependent zero-bias conductance $G_{\sigma\sigma'}$ 
in armchair graphene nanoribbons (AGNRs) with hydrogen adsorbates  employing a DFT-based 
\textit{ab initio} transport formalism including spin-orbit interaction (SOI). 
We find that a narrow AGNR decorated with a single hydrogen adatom
exhibits a spin-flip conductance $G_{\sigma\bar{\sigma}}$  that is highly anisotropic. 
In the case of collinear conducting and local impurity spins, the spin-flip conductance is due to SOI and it is very small, at most $10^{-4}\,e^2/h$.
In contrast, in the non-collinear situation, we observe a spin-flip conductance between 10$^{-2}\,e^2/h$ and 10$^{-1}\,e^2/h$, that can 
even exceed the spin-conserving conductance in some energy range. 
We explain this effect by an exchange-mediated spin-flip mechanism masking 
any potential spin-flip effect by spin-orbit interaction. 
Our calculations suggest that the spin-dependent conductance becomes isotropic again, 
if the concentration of adatoms is not too small. In this case the exchange mediated 
spin-flip scattering is always strong.

\section*{Acknowledgments} 
We express our gratitude to F.~Weigend for drawing our attention to the all-electron spin-orbit 
module~\cite{SOCMiddendorf} and helpful advice on it. We also thank  R.~Koryt\'{a}r, A.~Bagrets and I.~Beljakov for 
helpful discussions on spin DFT. The authors acknowledge the DFG (EV30/7-1 and EV30/8-1) for financial support 
and express their gratitude to the Simulation Lab NanoMicro, especially to I.~Kondov, for computational support. 
Computing time granted by the High Performance Computing Center Stuttgart (HLRS) on the HERMIT supercomputer, 
and computing time on the HC3 cluster at Karlsruhe Institute of Technology (KIT), operated by the 
Steinbuch Center for Computing (SCC), is gratefully acknowledged.

\appendix
\section*{Appendix: Toy model, computation of conductance and current density}
In App.~\ref{sec:app2}, we present details on the computation of the spin-dependent conductance.
In App.~\ref{sec:sec5}, we introduce a toy model to rationalize the shape and 
the order of magnitude of the spin-dependent conductance of the ribbon with
a single hydrogen adatom. 
Appendix~\ref{sec:currents} provides information 
on the formalism to calculate the current density shown in Fig.~\ref{fig:j}.

\section{Computation of the spin-dependent conductance}\label{sec:app2}
\subsection{Method details: Expansion of operators in real-space basis functions and partitioning in device and contact region}
In this section, we describe the expansion of all operators appearing in Eqs.\,\eqref{eq1}\,--\,\eqref{eq5} in basis functions of the underlying DFT calculation and how we partition the appearing matrices in blocks belonging to the device region and the contact regions to the reservoirs. 

From a non-periodic open-shell DFT calculation including all-electron spin-orbit coupling (SOI)~\cite{SOCMiddendorf}, we obtain the Kohn-Sham (KS) matrix $\mathbf{H}$ of a finite-size ribbon [e.g.~see Fig.\,\ref{fig2}\,(b)] with matrix elements 
\begin{align}
H^{\mu\nu}_{\sigma\sigma'}=\int \hspace{-0.05cm}d^3\mathbf{r}\,\varphi^\mu(\mathbf{r}) \ham_{\sigma\sigma'}\varphi^\nu(\mathbf{r})\,.
\end{align}
$\{\varphi^\nu\}_{\nu=1}^N$ denote the basis functions constructed from the underlying DFT calculation~\cite{def2SVP} which are real-valued, atom-centered and orthogonalized (via L\"owdin orthogonalization~\cite{Loewdin}).  $\mathbf{H}$ is of size $2N{\times}2N$ with non-zero entries in the off-diagonal spin-blocks due to SOI.  

Subsequently, we cut off every element~$H^{\mu\nu}_{\sigma\sigma'}$ from $\mathbf{H}$, if $\varphi^\nu$ and/or $\varphi^\mu$ is centered on an atom outside the device region, the latter indicated by the blue boxes in the previous figures. Employing the resulting truncated $2\tilde{N}{\times}2\tilde{N}$-device-KS matrix~$\mathbf{H}_\text{device}$ ($\tilde{N}\kt N$), we calculate the $2\tilde{N}{\times}2\tilde{N}$-matrix representation of the device-Green's function in presence of the left and the right reservoirs as
\begin{align}
\mathbf{G}(E) = \left[ E\,\mathds{1}_{2\tilde{N}}- \mathbf{H}_\text{device} - \boldsymbol{\Sigma}^L(E) - \boldsymbol{\Sigma}^R(E)     \right]^{-1}\,.
\end{align}
The $2\tilde{N}{\times}2\tilde{N}$-self-energy matrices~$\boldsymbol{\Sigma}^\alpha(E)$ are computed by a separate DFT calculation of a clean, finite-size and closed-shell treated AGNR11 ribbon, see  Ref.~\cite{MichaelsMethod} for details.  The self-energy matrices $\boldsymbol{\Sigma}^\alpha(E)$ are spin block-diagonal because of the closed-shell treatment of the reservoirs. Their non-vanishing entries correspond to the contact regions (marked by magenta boxes in Fig.~\ref{fig7}) which -- throughout this paper --were chosen as the outermost left and the outermost right column of carbon rings inside the (blue marked) device-region (i.\,e.~22 carbon atoms belong to each of the left and right contact region).

Then, we partition~$\mathbf{G}(E)$ into four $\tilde{N}{\times}\tilde{N}$-block matrices $\mathbf{G}_{\sigma\sigma'}(E)$ and $\boldsymbol{\Sigma}^\alpha(E)$ into two $\tilde{N}{\times}\tilde{N}$-block matrices $\boldsymbol{\Sigma}_{\sigma\sigma}^\alpha(E)$ [shorthand notation: $\boldsymbol{\Sigma}_{\sigma}^\alpha(E)$ with $\boldsymbol{\Sigma}_{\sigma}^\alpha(E)\eqt \boldsymbol{\Sigma}_{\bar{\sigma}}^\alpha(E)$ due to the closed-shell electronic structure of the reservoirs]. Finally, the spin-dependent conductance~$G_{\sigma\sigma'}(E)$ is computed as an orbital trace of a product of  $\tilde{N}{\times}\tilde{N}$-matrices,
\begin{align}
G_{\sigma\sigma'}(E) = \frac{e^2}{h}\,\text{Tr}\hspace{-0.1em}\left[\boldsymbol{\Gamma}^L_\sigma(E) \,(\mathbf{G}_{\sigma\sigma'}(E))^\dagger\,\boldsymbol{\Gamma}^R_{\sigma'}(E)\, \mathbf{G}_{\sigma'\sigma}(E) \right]\,,\label{eqb3}
\end{align}
where $\boldsymbol{\Gamma}_\sigma^\alpha(E)\eqt i(\boldsymbol{\Sigma}^\alpha_\sigma(E) \mt (\boldsymbol{\Sigma}^\alpha_\sigma(E))^\dagger)$.

\subsection{Validiation of the partitioning in device and contact region}\label{subsec:apppart}
\begin{figure}[t]
\centering
\includegraphics[width=8.6cm]{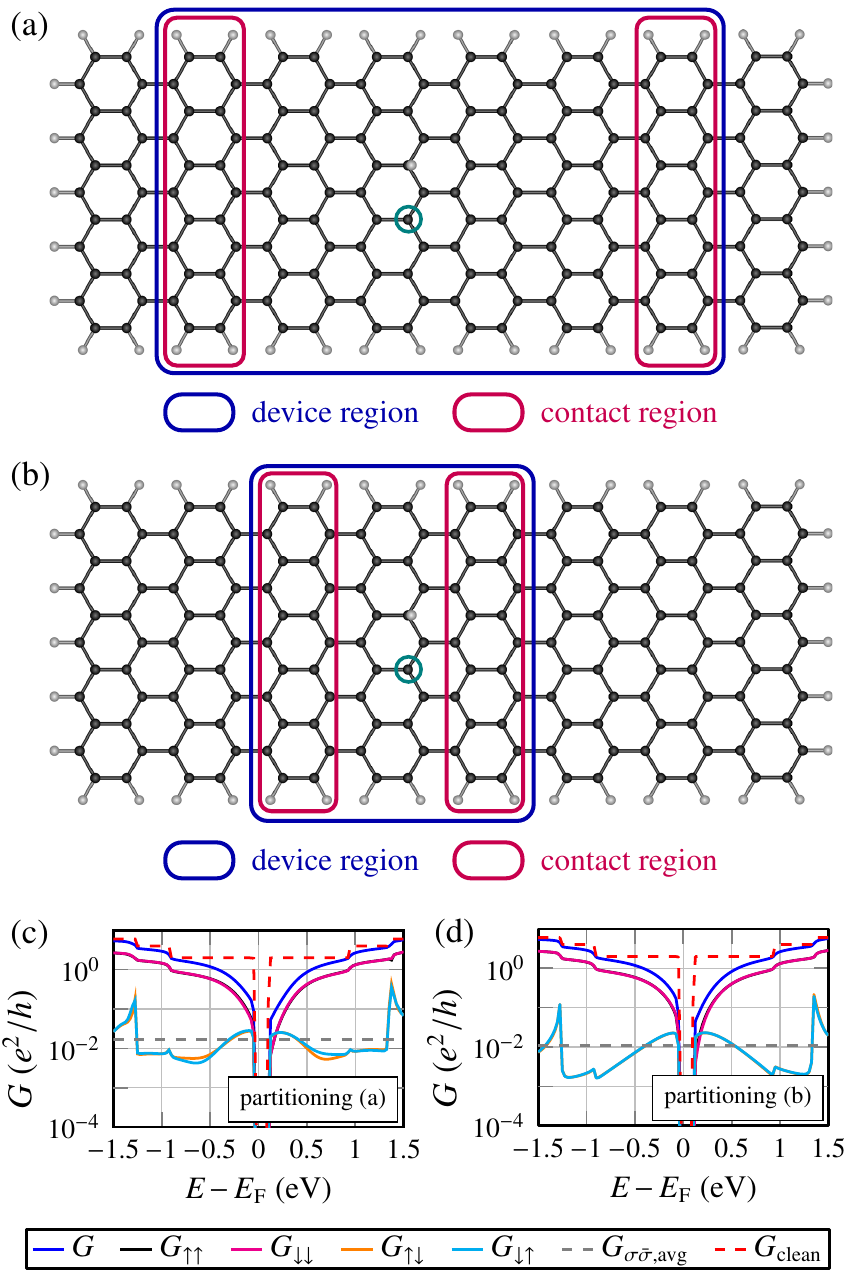}
\caption{
(a) Input geometry for the SOI-DFT calculation of an AGNR11 with a single adsorbed hydrogen atom.  The computed total magnetic moment of the ribbon is ${\braket{\mathbf{S}}}\eqt(-\,0.71,-\,0.48,-\,0.97)^T\hbar$, so that there are 2.7 unpaired electrons in the ribbon. For the transport calculation, we partition the ribbon into a device region (blue box) and two contact regions (magenta boxes). 
(b) Identical geometry and identical DFT calculation as in (a), but different partitioning in device and contact regions.
(c)/(d) Result for the spin-dependent conductance according to Eq.~\eqref{eqb3} for partitioning (a) [shown in (c)] and for partitioning (b) [shown in (d)]  for $z$-quantization so that ${\mathbf{e}_{\braket{\mathbf{S}}}}\hspace{-0.12em}\neqt \mathbf{e}_z$ enabling the efficient exchange-mediated spin-flip mechanism. 
}
\label{fig7}
\end{figure}

In this section, we show that the spin-dependent conductance is only weakly dependent on the precise (numerical) partitioning of the finite-size ribbon into device and contact region. 
As an example, we consider a ribbon with a single hydrogen adatom [as in Fig.\,\ref{fig2}\,(a)]. As finite-size input geometry for the SOI-DFT calculation, we choose a long finite-size ribbon, see Fig.\,\ref{fig7}\,(a)/(b), but we define two different device regions, as sketched by the blue boxes. The contact regions with non-vanishing self-energy are chosen as outermost row of carbon rings in the device region as indicated by the magenta boxes.

The computed conductance~$G_{\sigma\sigma'}(E)$ for the large device region of Fig.\,\ref{fig7}\,(a) is shown in Fig.\,\ref{fig7}\,(c) and  for the small device region of Fig.\,\ref{fig7}\,(b) in Fig.\,\ref{fig7}\,(d). The spin-conserving conductance~$G_{\sigma\sigma}(E)$ in (c) and (d) are indistinguishable from each other, while the spin-flip conductance~$G_{\sigma\bar{\sigma}}(E)$ in (c) and (d) exhibit the same shape. They are only deviating quantitatively from each other in energy intervals, where~$G_{\sigma\bar{\sigma}}(E)$ is anyway small. Additionally, the order of magnitude of~$G_{\sigma\bar{\sigma}}(E)$ agrees to the one in Fig.\,\ref{fig2}\,(d)/(f).
\pagebreak

As discussed in the body of this paper, the high spin-flip conductance is due to exchange with the magnetic moment of the device region. As can be seen in Fig.\,\ref{fig2}\,(b), the unpaired electrons are localized near the zigzag edges and the hydrogen impurity. For a calculation of~$G_{\sigma{\sigma'}}(E)$  without any finite-size artifacts, the full local moment near the hydrogen adsorbate has to be included in the device region, but no local moments originating from the zigzag edges. 
We conclude by comparing~$G_{\sigma\bar{\sigma}}(E)$  in Fig.\,\ref{fig7}\,(c) and (d), that local moments of the zigzag edge states are included in the large device region of Fig.\,\ref{fig7}\,(a) and/or that not the whole local moment caused by the hydrogen adsorbate is contained in the small device region of Fig.\,\ref{fig7}\,(b). However, the difference between~$G_{\sigma\bar{\sigma}}(E)$ of Fig.\,\ref{fig7}\,(c) and (d) is not serious as mentioned above. Consequences of changing the contact region are investigated in Ref.~\cite{MichaelsMethod}.

\section{Toy model for spin flips due to local exchange}\label{sec:sec5}
\begin{figure}[t]
\centering
\includegraphics[width=8.6cm]{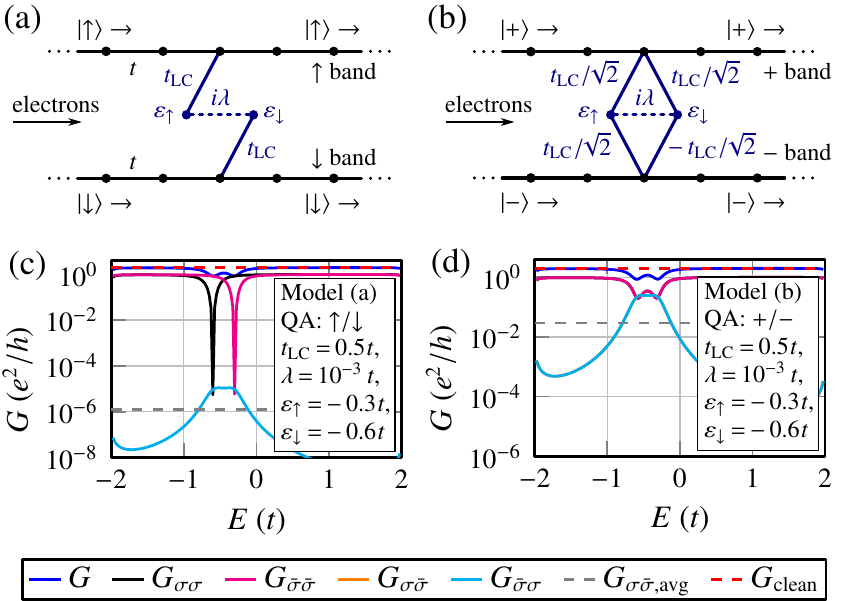}
\caption{(a)/(b) Tight-binding model to rationalize the conductance curves of the AGNR with a 
single hydrogen adsorbate including two conducting bands, a localized state with spin $\uparrow$ 
and $\downarrow$ and a SOI between both localized states. The matrix elements 
between the quantum wire  and the localized states are (a) $t_\text{LC}$ in the case of incoming 
electrons being spin-polarized along ${\uparrow}/{\downarrow}$ and (b) $t_\text{LC}/\hspace{-0.1em}\sqrt{2}$ for $+/-$-polarization. 
Here, one hopping term is negative, see Eq.~\eqref{e10}. 
(c)/(d) Spin-dependent conductance for model parameters for the AGNR with hydrogen 
adatom and (c) ${\uparrow}/{\downarrow}$- and (d) $+/-$-quanti\-zation of incoming electrons. The SOI 
strength $\lambda$ is estimated by the SOI strength of 2.5\,meV for a hydrogen 
adatom on graphene~\cite{SOCHydrGrapheneBala} in relation to the graphene hopping $t\eqt2.8$\,eV.
}
\label{fig3}
\end{figure}

To explain the shape and order of magnitude of $G_{\sigma\sigma'}(E)$ of Fig.~\ref{fig2}, 
consider a toy model that consists of the Hamiltonian of the lead and a single site
\begin{align}
\label{e5}
\ham_0 &= -\,t_0 \sum_{\sigma=\uparrow,\downarrow}\sum_{x}\left(\ch^\dagger_{x+1,\sigma}\ch_{x,\sigma}+\ch^\dagger_{x,\sigma}\ch_{x+1,\sigma}\right)  \\[0.3em]
\ham_\text{d}&=\sum_{\sigma=\uparrow,\downarrow}\varepsilon_{\sigma}\dhat^\dagger_{\sigma}\dhat_{\sigma} 
+ i\lambda \left(\dhat^\dagger_{\uparrow}\dhat_\downarrow - \dhat^\dagger_{\downarrow}\dhat_\uparrow\right) 
\label{e6}
\end{align}
and a coupling term 
\begin{equation}
\label{e7}
\pot = t_\text{LC}\sum_{\sigma=\uparrow,\downarrow} \left(\dhat^\dagger_{\sigma}\ch_\sigma + \ch^\dagger_{\sigma}\dhat_\sigma\right)
\end{equation}
with $\hat{c}_\sigma\equivt\hat{c}_{0,\sigma}$. The whole model is sketched in Fig.\,\ref{fig3}\,(a). It resembles a hydrogen adsorbate that splits off a resonant 
level~\cite{LiLu,ElPropGr,*MidgapStatesPRL,*MidgapStatesPRLFe}  near the charge-neutral point from the conduction band continuum for both spin 
channels. The exchange interaction 
with the local spin lifts the degeneracy between the localized states with differing 
spin: $\varepsilon_{\uparrow}\neqt\varepsilon_{\downarrow}$. In the case of a non-vanishing 
overlap matrix element $t_\text{LC}$, the associated quasilocalized state contributes a 
separate conductance channel that interferes destructively with the 
residual ones~\cite{Antires}.    The SOI is modelled to be only present at the 
resonant site by the parameter~$\lambda$. 
%%%
The model Eqs.\,\eqref{e5}\,--\,\eqref{e7} account for collinear 
conducting and local spins reflecting in the vanishing overlap between $\sigma$ 
band and $\bar{\sigma}$ quasilocalized state. The spin-dependent conductance of 
the collinear-spin model of Fig.\,\ref{fig3}\,(a) is depicted in (c).  We observe 
the well-known conductance dips at the resonant energies $\varepsilon_{\uparrow}$ 
and $\varepsilon_{\downarrow}$~\cite{LiLu} and a small spin-flip conductance in the 
order of $(\lambda/t)^2$ due to the small SOI parameter~$\lambda$.

%%%%%%%%%%%%%%%%%%%%%%%%%%%%%%%%%%%%%%%%%%%%%%%%%%%%%%%%%%%%%%%%%%%%%%%%%%%%%%%%%%%%%%%%%%%%%%%%%%%%%%%%%%%%%%%
When rotating the spin quantization of incoming electrons by $\pi/2$ from ${\uparrow}/{\downarrow}$ to $+/-$, 
$
{\ket{+}} \eqt ({\ket{\uparrow}}\pt{\ket{\downarrow}})/\hspace{-0.1em}\sqrt{2}
$
and
$
{\ket{-}} \eqt ({\ket{\uparrow}}\mt{\ket{\downarrow}})/\hspace{-0.1em}\sqrt{2}
$,
an overlap of both quasilocalized states to both bands is formed. To account for this we generalize our model 
\begin{align}
\label{e8}\ham_0 &= -\,t_0 \sum_{\mu=+,-}\sum_{x}\left(\ch^\dagger_{x+1,\mu}\ch_{x,\mu}+\ch^\dagger_{x,\mu}\ch_{x+1,\mu}\right)  \\[0.3em]
\ham_\text{d}&=\sum_{\sigma=\uparrow,\downarrow}\varepsilon_{\sigma} \dhat^\dagger_{\sigma}\dhat_{\sigma} 
+ i\lambda \left(\dhat^\dagger_{\uparrow}\dhat_\downarrow - \dhat^\dagger_{\downarrow}\dhat_\uparrow\right) 
\label{e9}
\end{align}
and the coupling 
\begin{equation}
\label{e10}
\pot = \frac{t_\text{LC}}{\sqrt{2}}\left( \dhat^\dagger_{\uparrow}\left(\ch_++\ch_{-}\right) 
+ \dhat^\dagger_{\downarrow}(\ch_{+}-\ch_{-}) + \text{h.c.}\right)
\end{equation}
where $\ch_{+}\eqt(\ch_\uparrow\pt\ch_\downarrow)/\hspace{-0.1em}\sqrt{2}$ and $\ch_{-}\eqt(\ch_\uparrow\mt\ch_\downarrow)/\hspace{-0.1em}\sqrt{2}$
[model in Fig.\,\ref{fig3}\,(b)]. It now imitates the situation where the local moment and the 
conduction band spin are not collinear. Then, even in the absence of SOI, $\lambda\eqt0$, 
the conduction band spin is not conserved and we expect a large spin-flip rate. 
Indeed, seen in Fig.\,\ref{fig3}\,(d),  
the spin-flip conductance increases by four orders of magnitude compared to (c) 
even exceeding the spin-conserving conductance near the antiresonance. 
%%%
The results of the model agree well with the conductance of the ribbon with 
a single hydrogen adatom in Fig.~\ref{fig2}: 
In particular, it also reproduces the quasilocalized state also seen in or near the bandgap of the 
ribbon.

We mention that in the non-magnetic case, $\varepsilon_{\uparrow}\eqt\varepsilon_{\downarrow}$, 
the efficient spin-flip process for $+/-$-polarization is suppressed. The reason is that in this case the single site effectively 
acts like a two-path interferometer that support perfect destructive interference for the 
two tunneling paths between the $+$ and the $-$~band. 
\\

\section{Computation of the local current density}\label{sec:currents}
For completeness, we summarize the calculation of the local current density response $\vv j(\vv r, E)$ starting from the full spin-dependent retarded Green's function $\hat G(E)$ [see Eq.~\eqref{eq1} for its spin components]. The retarded Green's function allows to calculate the non-equilibrium Keldysh Green's function~$\hat G^{<}(E)$:\vspace{-1em}
\begin{align}
  \hat G^{<} = i \hat G \big[f^\text{L} \hat\Gamma^\text{L}+f^\text{R} \hat\Gamma^\text{R}\big] \hat G\hc\mathcomma
\end{align}
with $\hat{\Gamma}^\alpha\eqt i(\se^\alpha \mt (\se^\alpha)^\dagger)$. The occupation numbers $f^\alpha$ of the leads reduce to step functions at zero (or low) temperature. Inside the voltage window, we assume an occupied left lead and an empty right lead, i.e., $f^\text{L}\eqt1$, $f^\text{R}\eqt 0$, so that $\hat G^<$ reduces to\vspace{-1em}
\begin{align}
  \hat G^<(E) = i \hat G(E) \hat \Gamma^\text{L}(E) \hat G\hc(E)\mathperiod
\end{align}
The Keldysh Green's function is transformed to real-space representation using 
the basis functions of the underlying DFT calculation: $G_{\sigma\sigma'}^<(\vv r, \vv r', E) \eqt \langle{\vv r\sigma}|{\hat G^<(E)}|{\vv r'\sigma'}\rangle$. The current density (per energy) is then expressed as 
\begin{equation}\label{eq:currentDensity}
  \vv{j}(\vv r, E) = \frac{1}{2\pi} \frac{\hbar}{2m} \sum_{\sigma}\,\lim\limits_{\vv r'\to \vv
r} (\VV\nabla_{\vv r'} - \nablagrad_{\vv r}) G_{\sigma\sigma}^<(\vv r, \vv r',
E)\mathperiod
\end{equation}
The factor $2\pi$ reflects an inverse Fourier transform.

\bibliography{Literature}

\end{document}